\documentclass{article}
\usepackage{ais25}
\usepackage[utf8]{inputenc} 
\usepackage[T1]{fontenc}    
\usepackage{hyperref}       
\usepackage{url}            
\usepackage{booktabs}       
\usepackage{amsfonts}       
\usepackage{nicefrac}       
\usepackage{microtype}      
\usepackage{xcolor}         
\usepackage{tablefootnote}

\usepackage{comment} 
\usepackage{todonotes} 

\usepackage{graphicx}
\usepackage{subfigure}

\title{Exploring Image–Text Alignment for Radio Galaxy Morphologies}

\author{\name{Erica Lastufka, Mariia Drozdova, \and Svyatoslav Voloshynovskiy} \\ \addr{University of Geneva}  \email{\{erica.lastufka,mariia.drozdova,svyatoslav.voloshynovskyy\}@unige.ch} \\ 
}

\begin{document}
\maketitle

\begin{abstract}
We investigate whether specially constructed text captions can capture the same morphological information as radio galaxy images. Using the MiraBest dataset, we generate captions with a domain-specific prompt and evaluate their alignment with images through the SigLIP-2 vision--language model, with and without LoRA fine-tuning. Results show that caption-based classification of FR-I and FR-II galaxies performs similarly to images, with fine-tuning improving local coherence of embeddings but not global alignment. 
\end{abstract}



\section{Introduction}

Vision--language models (VLMs) have demonstrated remarkable zero-shot generalization in natural image domains, learning joint embeddings of images and text that enable retrieval, captioning, and reasoning tasks. In astrophysics, the direct application of these models for scientific purposes is not straightforward, as text is not a common data format in either observations or simulations. While specialized large language models such as \textit{AstroMLab} \citep{pan_astromlab_2024} and \textit{AstroLLaMA} \cite{pan_astromlab_2024} capture domain-specific textual information from literature and metadata, and self-supervised visual models such as \citet{lastufka_self-supervised_2024, hayat_self-supervised_2021, slijepcevic_radio_2024} extract  features from images, no studies to our knowledge investigate whether text representations can encode the same morphological information as astrophysical image data. 

Recent work in this direction includes \textit{AstroLLaVA} \cite{zaman_astrollava_2025}, which aligns captions with pictures from NASA's Astronomy Photo of the Day, and CosmoCLIP \citep{imam_cosmoclip_2024}, which uses BLIP-generated captions together with scraped pictures for contrastive fine-tuning.
Although these works feature galaxies, planets, and astrophysical objects, they do not use scientific images; that is, the type of image that a scientist would generally analyze. The generated or written captions also reflect the popular audience, and contain global descriptions rather than detailed morphological information. As an intermediate step between catalog- or meta-data and images, perhaps natural language descriptions of source morphology can enable commercial LLMs to be leveraged for scientific tasks.

Natural language descriptions obtained in this way might complement existing source catalogs, which are usually produced by fitting Gaussians to flux values above certain thresholds. Not all features of extended, non point-source emission can be captured in a catalog, so a description such as "an asymmetrical, diffuse lobe extending diagonally" might contain more semantic information than tabular catalog data of individual, non-overlapping Gaussians\footnote{blended or overlapping sources are often omitted from source catalogs if they cannot be sufficiently distinguished}. Therefore, natural language captions, should they prove informative, could be an label of global image properties, as opposed to the patch-level properties captured by source catalog measurements.
Furthermore, the use of VLM-generated captions tests an interesting theoretical concept, that of the information bottleneck \citep{tishby_information_2000}. The idea that a model encodes task-relevant information from an images, and then decodes it into the language modality, would lend insight into how a model trained on natural images represents scientific knowledge.  

In this work, we compare the captioning quality resulting from a specialized prompt vs a generic prompt, and we investigate whether these synthetic text representations can capture the same semantic information as images. We further explore lightweight adaptation through LoRA fine-tuning of the vision and/or language backbones, examining how alignment between image and text embeddings can be improved. By systematically evaluating similarity and classification scores, we aim to quantify the information preserved in text embeddings and eventually to assess their potential to enhance zero-shot capabilities of vision models in astrophysical applications.

\section{Data}

Experiments use the small, curated MiraBest dataset \cite{porter_mirabest_2023}, which contains a roughly even split between Fanaroff--Riley type I (FR-I) and type II (FR-II) morphologies. Images are sparse, featuring small areas of bright to faint emission. The small sample is important for the ability to review the quality of caption generation. Additionally, this dataset has been used in prior studies, such as \citet{mohale_enabling_2024, slijepcevic_radio_2024, lastufka_self-supervised_2024, riggi_evaluating_2025} enabling straightforward comparison. MiraBest sources are labeled as \textit{Confident} or \textit{Uncertain} according to the majority vote of human annotators'  morphological classification. For training, we exclude Uncertain samples to reduce noise, resulting in 833 Confident images. A representative subset of 104 Confident samples is held out as a test set. 

To generate image captions, we employ Google's proprietary Gemini 2.5-Flash preview-05-20 model, using a domain-specific prompt designed to emulate expert morphological analysis. A control prompt simply requests generation of captions describing the images, while the experimental prompt asks for a detailed morphological description\footnote{for full prompts, see Appendix \ref{sec:prompts}}. Control captions exhibited a larger ($\sim$1.5x) vocabulary than those in the experimental dataset, although they had less morphological detail and often assumed incorrect descriptors.
Captions produced by the experimental prompt, henceforth referred to as the curated captions, were manually controlled, and if need be, edited for the test set only. 

Generally, curated captions were found to describe the key features of the radio sources, without injecting too much interpretation. 
Many captions used similar wording, but given the similarity of some of the images, this was likely unavoidable due the limitations of the prompt. Overall, the captions were found to be detailed and objective, similar to what a scientist might compose.\footnote{examples of images and captions are in Appendix \ref{sec:examples}} 

\section{Experiments and Results}

We evaluated binary classification and similarity metrics using the open-source SigLIP-2 ViT-Base/patch-16 model with 224$\times$224 resolution. 
SigLIP-2 uses sigmoid loss, which treats image-text matching as a binary classification problem, together with location-aware captioners (LocCa), self-distillation and masked prediction. For our LoRA fine-tuning, we evaluate only the sigmoid loss, as the full SigLIP-2 loss function is not yet implemented. 
We evaluate models using F1 score from linear probes and $k$-nearest neighbors (KNN) classifiers\footnote{We chose k=5} on frozen embeddings, as well as embedding similarity metrics (Recall@1, Top-5 Recall, and class-level Recall@1). These metrics allow us to examine the separability of embeddings for classification tasks, as well as the alignment of image and text latent spaces\footnote{for full definition and interpretation of chosen metrics, see Appendix \ref{sec:eval}}.

Parameter-efficient fine-tuning was performed using LoRA \citep{hu_lora_2021}. We experimented with fine-tuning only the vision encoder ("Vision" column, 0.17\% of total parameters trainable), only the text encoder ("Text" column 0.16\% of total parameters trainable), and both vision and text encoders ("Full" column, 0.33\% of total parameters trainable). LoRA parameters of $R$=8 and $\alpha$ = 16 were used, together with a learning rate of 2.5e-5. Training was performed for a maximum of 50 epochs, with early stopping implemented using loss calculated on a small validation subset of the training set ($\sim$100 samples). 
Results are displayed in Table \ref{tbl:results}, with the difference in performance between using the curated and control captions displayed in parentheses when applicable, and the best results highlighted in bold text.

\begin{table}[h]
\caption{Evaluation results}
\label{tbl:results}
\begin{center}
\begin{tabular}{p{5cm}p{1.25cm}p{1.25cm}p{1.25cm}p{1.25cm}}
\toprule
\textbf{Metric} & \textbf{Frozen}  & \textbf{Vision}  & \textbf{Text}  &\textbf{Full}  \\
     \midrule
Linear probe F1, images & 0.90 (-0.01) & 0.93 (+0.03) & 0.89 (-0.01) & \textbf{0.93 (+0.05)} \\
Linear probe F1, text & 0.90 (+0.01) & \textbf{0.92 (+0.03)} & 0.89  & \textbf{0.92 (+0.03)} \\
Linear probe F1, concatenated & 0.88 (+0.01) & 0.88 (-0.01) & 0.87 & \textbf{0.88 (+0.02)} \\
Mean image-text cosine similarity & \textbf{0.14}  & 0.13 (-0.01) & 0.13 (-0.01) & 0.13  \\
Recall@1 & 0.01 (-0.03) & 0.03 (-0.08) & 0.03  & \textbf{0.07 ($-$0.01)} \\
Top-5 Recall & 0.06 (-0.12) & 0.13 (-0.23) & \textbf{0.23 (+0.04)} & 0.23 (-0.18) \\
Class-level Recall@1 & 0.53 (-0.05) & 0.66 (-0.12) & 0.69 (+0.11) & \textbf{0.77}  \\
KNN-classifier F1, images & 0.81  & 0.86  & 0.81  & \textbf{0.88 (+0.03)} \\
KNN-classifier F1, text & 0.69 (-0.02) & 0.69 (-0.02) & \textbf{0.73 (+0.02)} & 0.70 (-0.08) \\
KNN-classifier F1, concatenated & 0.79 & \textbf{0.83 ($-$0.04)} & 0.74 (-0.05) & 0.81 (-0.04) \\
\midrule
\bottomrule
\end{tabular}
\end{center}
\end{table}


\section{Discussion and Conclusions}

\begin{figure}
    \centering
    \includegraphics[width=0.595\linewidth]{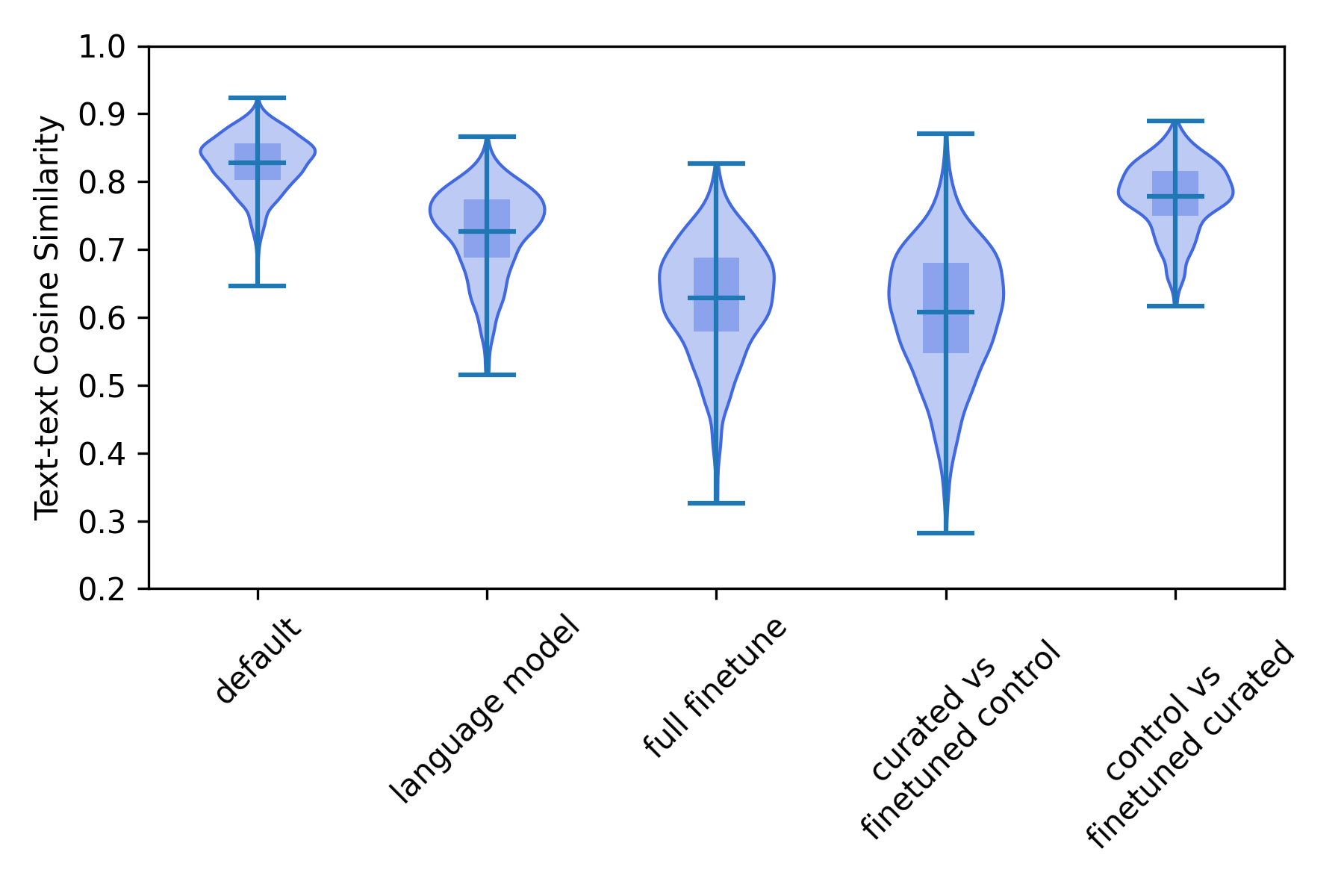}
    \includegraphics[width=0.395\linewidth]{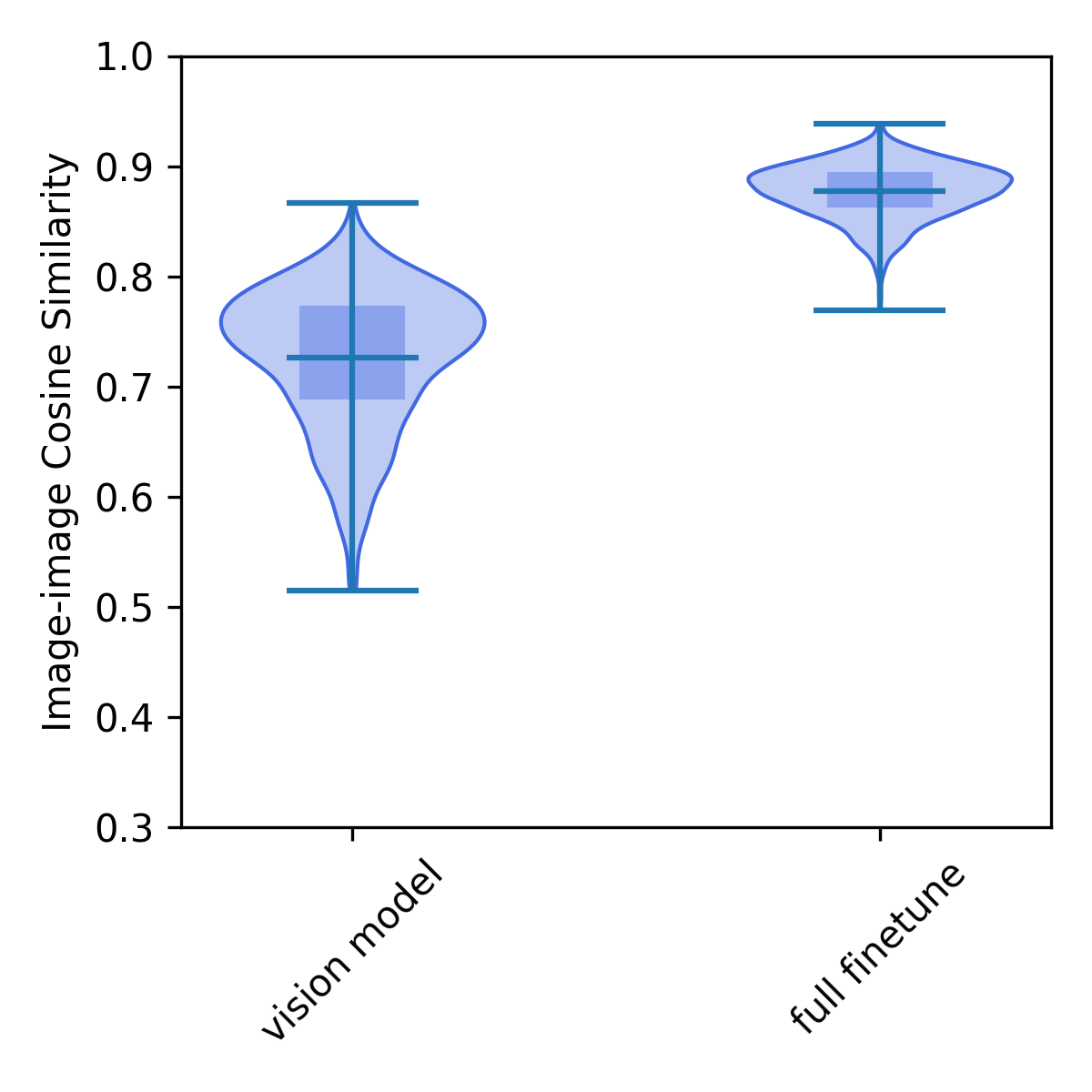}
    \caption{Left: Cosine similarity between the text embeddings of the curated and control captions, at different stages of model fine-tuning. Right: Cosine similarity between the image embeddings of the models jointly trained with the curated and control captions.
    }
    \label{fig:modality_cosine_sim}
\end{figure}

Linear probe metrics confirm that both images and text carry sufficient information to separate FR-I and FR-II galaxies. "Scientific" captions only provide a small boost in performance relative to generic (and sometimes incorrect) ones. 
This surprising result suggests that Gemini, even with a simplistic prompt, samples vocabulary that is already highly distinctive.
Figure \ref{fig:modality_cosine_sim} shows that the cosine similarity between curated and control captions is very high, and decreases with fine-tuning.
The similarity scores between the original control captions and fine-tuned curated captions reveal an asymmetrical semantic anchoring; the similarity is higher than the opposite situation, demonstrating that instead of fine-tuning resulting in more specialized "scientific" captions, it is in fact doing the opposite: the curated embeddings are being pulled toward the model's more generalist domain defined by the control captions.

Nevertheless, LoRA fine-tuning improves discrimination between classes. Fine-tuning of the text encoder alone did not impact performance as much as fine-tuning the vision encoder or the full model. This reflects the interpretation that fine-tuning of the text encoder leads to more generalized representations, which do not sufficiently reinforce the sometimes subtle differences between galaxy morphologies. 
Interestingly, although validation metrics showed higher F1 scores when using the concatenated image and text features, test metrics did not show any improvement. Given the small dataset size, this likely reflects overfitting due to the high dimensionality of the concatenated features, which adds noise rather than complementary signal to the linear probe.

The cosine similarity between text and images (see Appendix  \ref{sec:cosine_sim}, Figure \ref{fig:cosine_sim}) does not change on average after fine-tuning, despite the sigmoid loss, which computes similarity within a mini-batch, decreasing. This indicates that the model successfully reduced the loss without improving the global alignment of positive image-text pairs, which must then be achieved by pushing negative samples away, leading to sharper class boundaries and denser distributions in the latent space. This explains the improved F1 scores and the increased local coherence seen in the T-SNE plots (Appendix \ref{sec:tsne}), confirming that fine-tuning improves class-level structure even if global alignment remains limited.
Finally, the KNN-classifier results demonstrate how samples are grouped in latent space. Fine-tuning the full model or just the text model led to a noticeable improvement, indicating that embeddings became locally more coherent, improving class-level structure.

To continue this study, we have adapted the method of \citet{jose_dinov2_2024}, which fully trains additional transformer blocks appended to the frozen vision encoder, for use with our pre-trained SigLIP-2 backbone, and are testing the effectiveness of this setup. Additionally, we wish to explore a model with a generative component to examine linguistically how captions evolve.

This study concludes that caption-based embeddings can encode meaningful morphological information from radio galaxies and are a viable complement to image-based representations. However, although using more "scientific" captions leads to marginal performance improvement in classification, instance-level distinction is sacrificed, and fine-tuning can generalize rather than specialize the language embeddings. Despite this trade-off, the curated captions with specialized, low-variance vocabulary demonstrate the semantic control and structural coherence required to reliably encode subtle features, such as image contamination or diffuse emission. 
Overall, SigLIP-2's ability to encode high-level semantic information via text validates the idea that natural language captions can serve as an essential qualitative layer which complements the more patch-level labels currently available through astronomical source catalogs. This opens new avenues for multimodal research and applications in radio astronomy, as it provides a low-resource method of labeling survey datasets.

\newpage
\bibliography{VLM}

\newpage
\begin{appendix}

\section{Caption-generating Prompts}\label{sec:prompts}

For the control captions, the following prompt was used:

\begin{quote}
"Please use 50 words or less to write a caption describing this image."
\end{quote}

For the curated captions, the following prompt was used:

\begin{quote}
    "You are acting as an expert observational radio astronomer. Examine the following image and describe it in a concise manner. Look for bright centroids of flux and where in the image (left, right, top, bottom, center) they are located. Are there any lobes or jets associated with the flux centroids? From where do they originate and where do the end? Are there any sources in the image that are not associated with the central galaxy? Only describe morphology, leaving out any assumptions about the nature of the emission. Use 50 words or less."
\end{quote}

\section{Dataset Examples}\label{sec:examples}

\begin{figure}[h]
    \centering
    \includegraphics[width=0.24\linewidth]{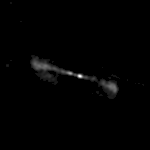}
    \includegraphics[width=0.24\linewidth]{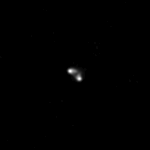}
    \includegraphics[width=0.24\linewidth]{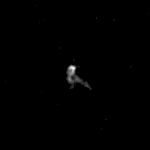}
    \includegraphics[width=0.24\linewidth]{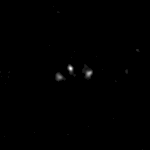}
    \caption{Examples of FR-I, FR-II, FR-I, and FR-II galaxies from the MiraBest training dataset. Captions are listed below.}
    \label{fig:examples}
\end{figure}

Figure \label{fig:examples} shows samples from the MiraBest dataset. Below, we list both the control and curated captions. Inaccuracies are highlighted in bold text, and irrelevant text in italics.

\subsection{Left image, FR-I}

Control caption:

\begin{quote}
This radio image shows a powerful radio galaxy with a bright central active galactic nucleus (AGN). It emits two opposing jets of plasma that expand into large radio lobes, extending far into space.
\end{quote}

Curated caption:

\begin{quote}
A bright, compact central source emits two opposing, collimated structures. These jets extend horizontally, broadening into diffuse, extended lobes terminating on the far left and right. No other distinct, unassociated sources are present.
\end{quote}

\subsection{Second image, FR-II}

Control caption:

\begin{quote}
\textbf{Three} bright, blurry objects, seemingly connected, are visible against a dark background. \textit{This image could depict distant celestial bodies or an out-of-focus view of faint lights in space.}
\end{quote}

Curated caption:

\begin{quote}
Two bright, closely spaced flux centroids are seen slightly top-right of center. They are surrounded by a fainter, diffuse, and irregular emission. Distinct lobes or collimated jets, with clear origins and ends, are not discernible from these centroids. No other significant radio sources are present.
\end{quote}

\subsection{Third image, FR-I}

Control caption:

\begin{quote}
A distant, irregularly shaped celestial body, \textbf{possibly an asteroid or comet}, appears as a blurry white object against the dark background of space. Its two-lobed structure suggests it might be a \textbf{contact} binary.
\end{quote}

Curated caption:

\begin{quote}
A bright, irregular source is slightly above image center. Its dominant flux centroid is upper-left, from which an elongated structure extends diagonally to the lower-right, blending into the source's diffuse emission. A fainter centroid lies within this extension. Several faint point sources are visible across the field.
\end{quote}

\subsection{Right image, FR-II}

Control caption:

\begin{quote}
Soft, blurry white glows emerge from a vast dark background. \textit{The clustered lights evoke deep space, distant phenomena, or abstract energy sources, creating an enigmatic view.}
\end{quote}

Curated caption:

\begin{quote}
Multiple bright flux centroids appear clustered slightly left of center, with two possibly blended. Another component is above and right, and a fainter one is to the far right. No distinct large-scale lobes or collimated jets are evident. Several very faint, unresolved point sources are scattered throughout.
\end{quote}


\section{Evaluation Metrics}\label{sec:eval}

The F1 score is the harmonic mean of precision and recall.

\textbf{Linear Probe F1 Score:}
A linear probe is a lightweight classifier trained on frozen embeddings, which tests whether embeddings preserve sufficient morphological information to distinguish between FR-I and FR-II galaxies.

\textbf{KNN Classifier F1 Score:}
A $k$-nearest neighbors (KNN) classifier assigns labels based on the majority class of the $k$ closest points in the embedding space. Unlike linear probes, KNN performance reflects local clustering properties of embeddings. Strong KNN F1 suggests that similar sources occupy compact neighborhoods, which is desirable for similarity-based discovery.

\textbf{Cosine Similarity:}
Cosine similarity measures the average alignment between paired image and caption embeddings. Higher similarity values indicate that the model places corresponding images and captions closer in the joint embedding space. 

\textbf{Recall@1 and Top-5 Recall:}
Retrieval metrics evaluate whether an embedding from one modality can identify its paired counterpart in the other modality. Recall@1 reports the fraction of cases where the correct caption is the top-ranked match for an image (or vice versa), while Top-5 Recall extends this to the top five matches. 

\textbf{Class-level Recall@1:}
This evaluates retrieval at the class level rather than the instance level. A prediction is counted as correct if the retrieved sample belongs to the correct morphological class, even if it is not the exact paired caption or image. Higher class-level recall suggests that text embeddings encode morphology sufficiently to cluster FR-I and FR-II galaxies, even if fine-grained instance-level alignment is lacking.

\section{Image-Text Cosine Similarity}\label{sec:cosine_sim}

\begin{figure}[h]
    \centering
    \includegraphics[width=\linewidth]{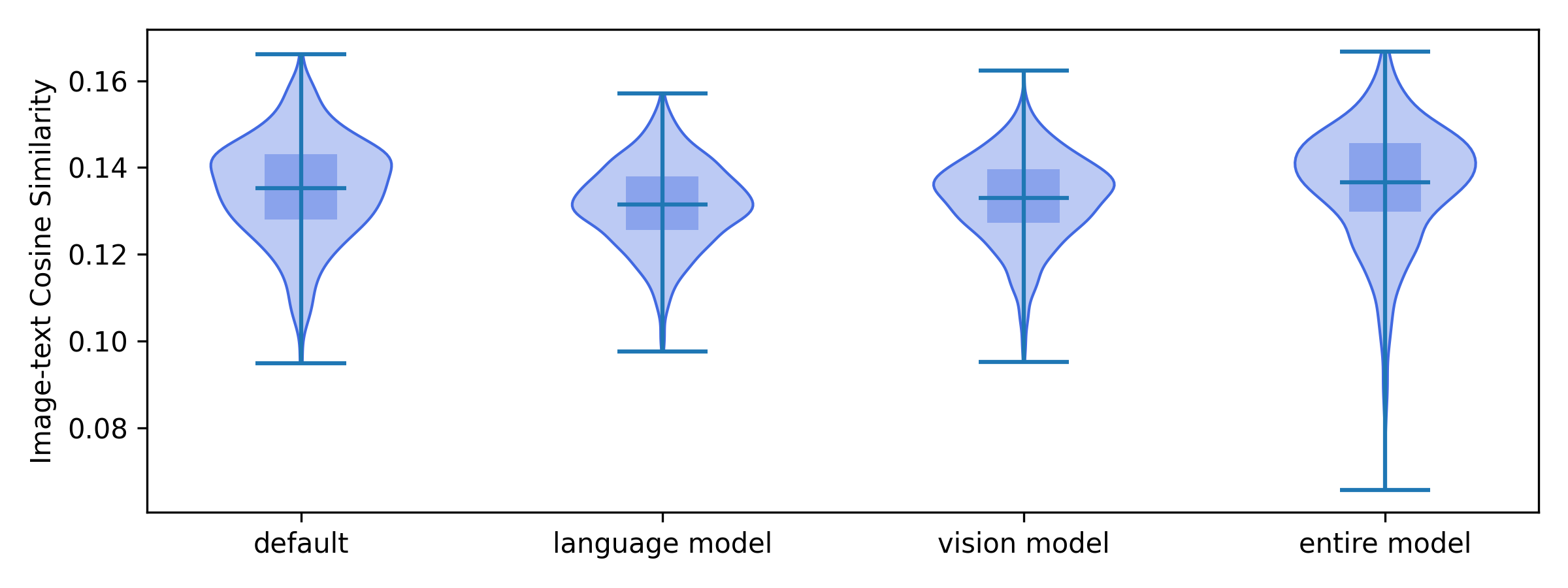}
    \caption{Cosine similarity between image and curated text embeddings for the MiraBest dataset. 
    }
    \label{fig:cosine_sim}
\end{figure}

\section{TSNE Latent Space Visualization}\label{sec:tsne}

\begin{figure}[h]
    \centering
    \includegraphics[width=0.327\linewidth]{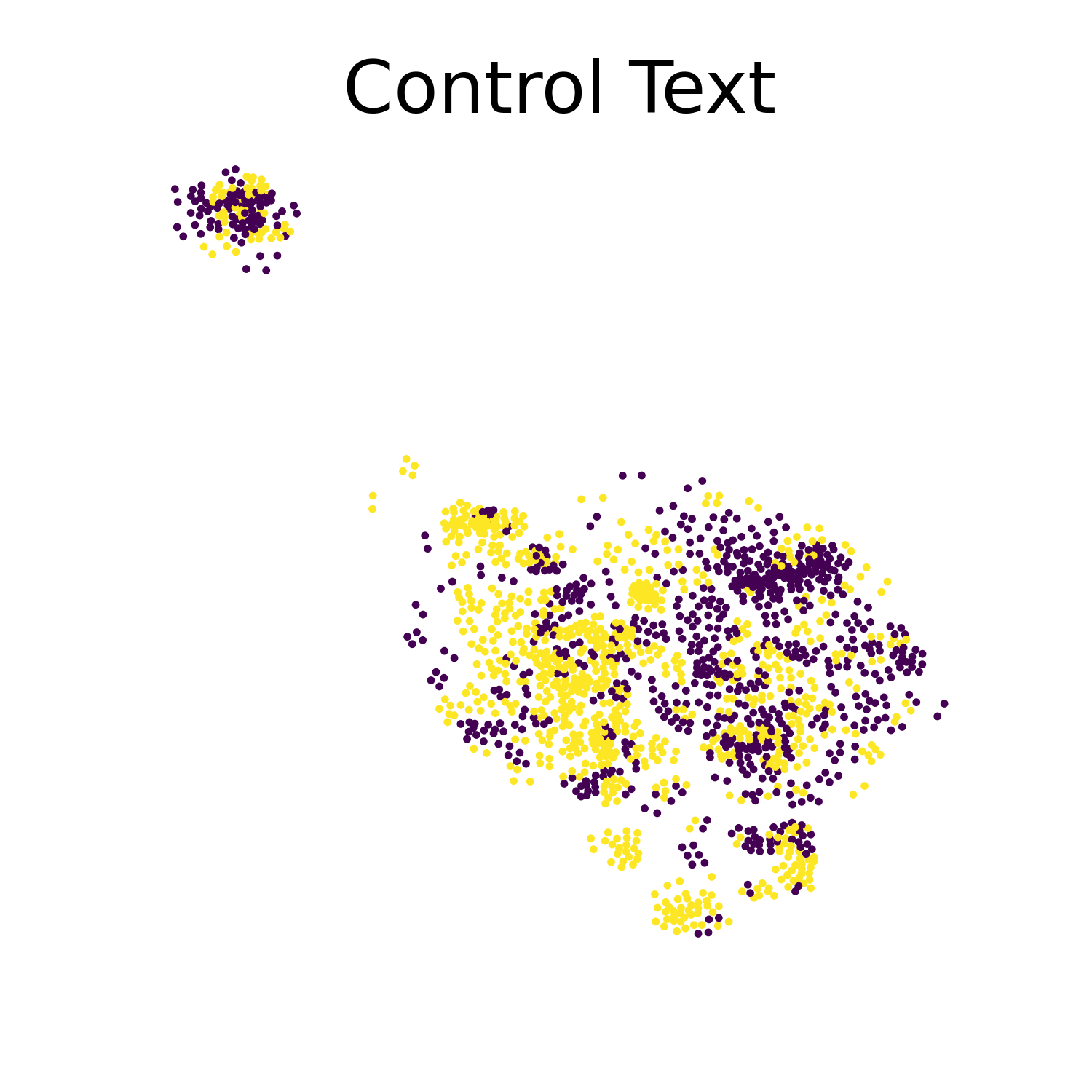}
    \includegraphics[width=0.327\linewidth]{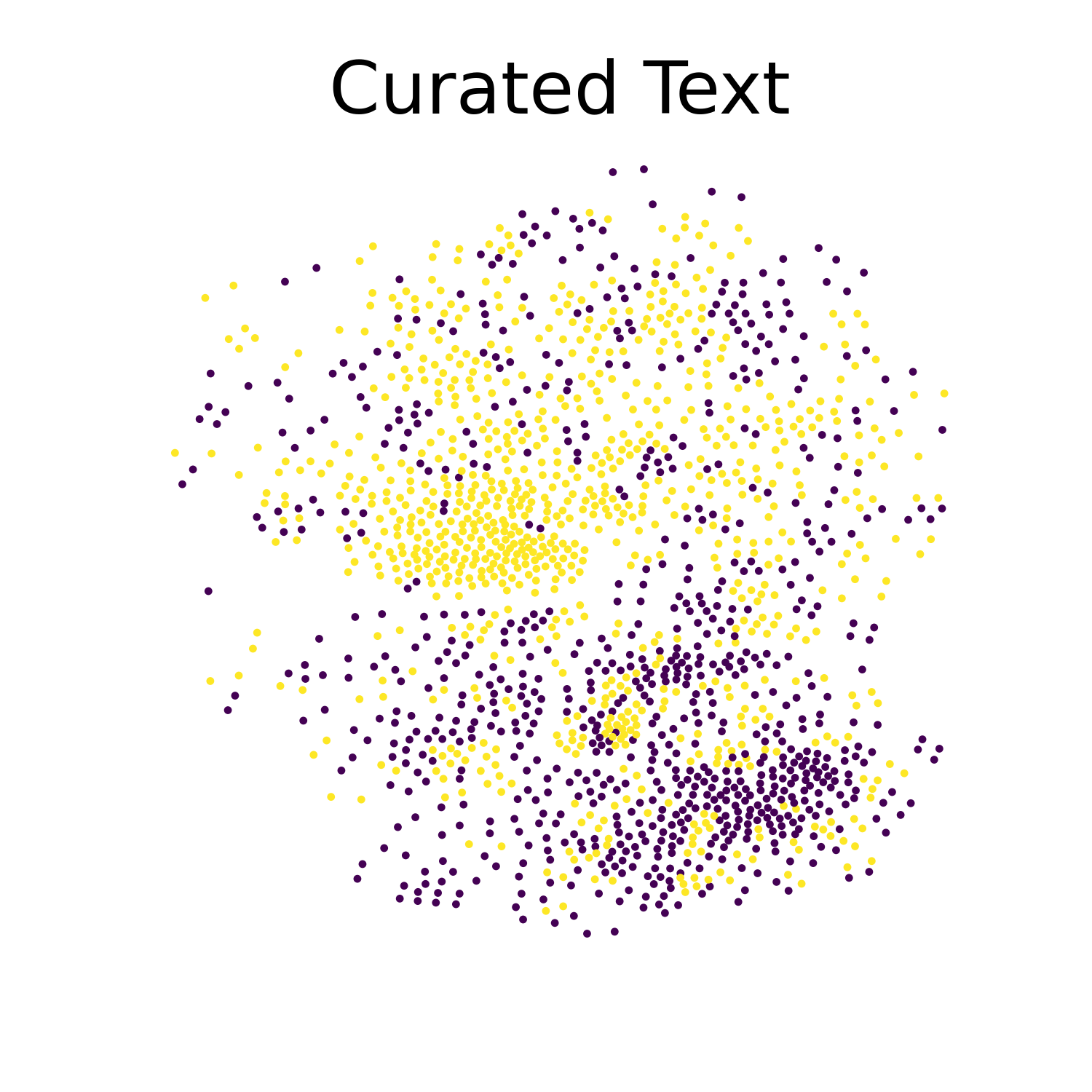}
    \includegraphics[width=0.327\linewidth]{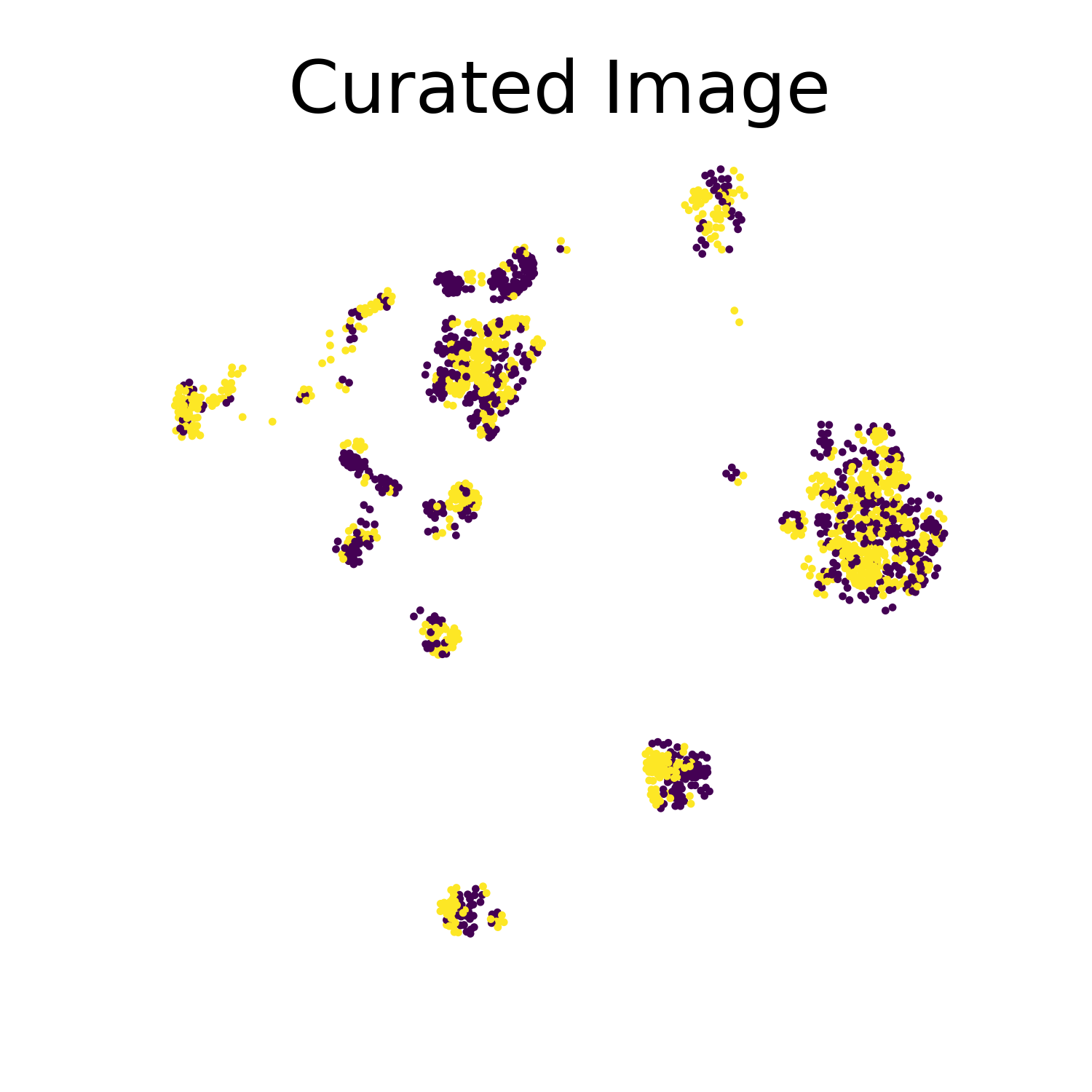}
    \caption{T-SNE visualization of text or features extracted from the frozen model. Image features are the same for both caption datasets. Morphology class FR-I is shown in purple, while FR-II galaxies are labeled in yellow.}
    \label{fig:tsne_frozen}
\end{figure}

\begin{figure}[h]
    \centering
    \includegraphics[width=0.24\linewidth]{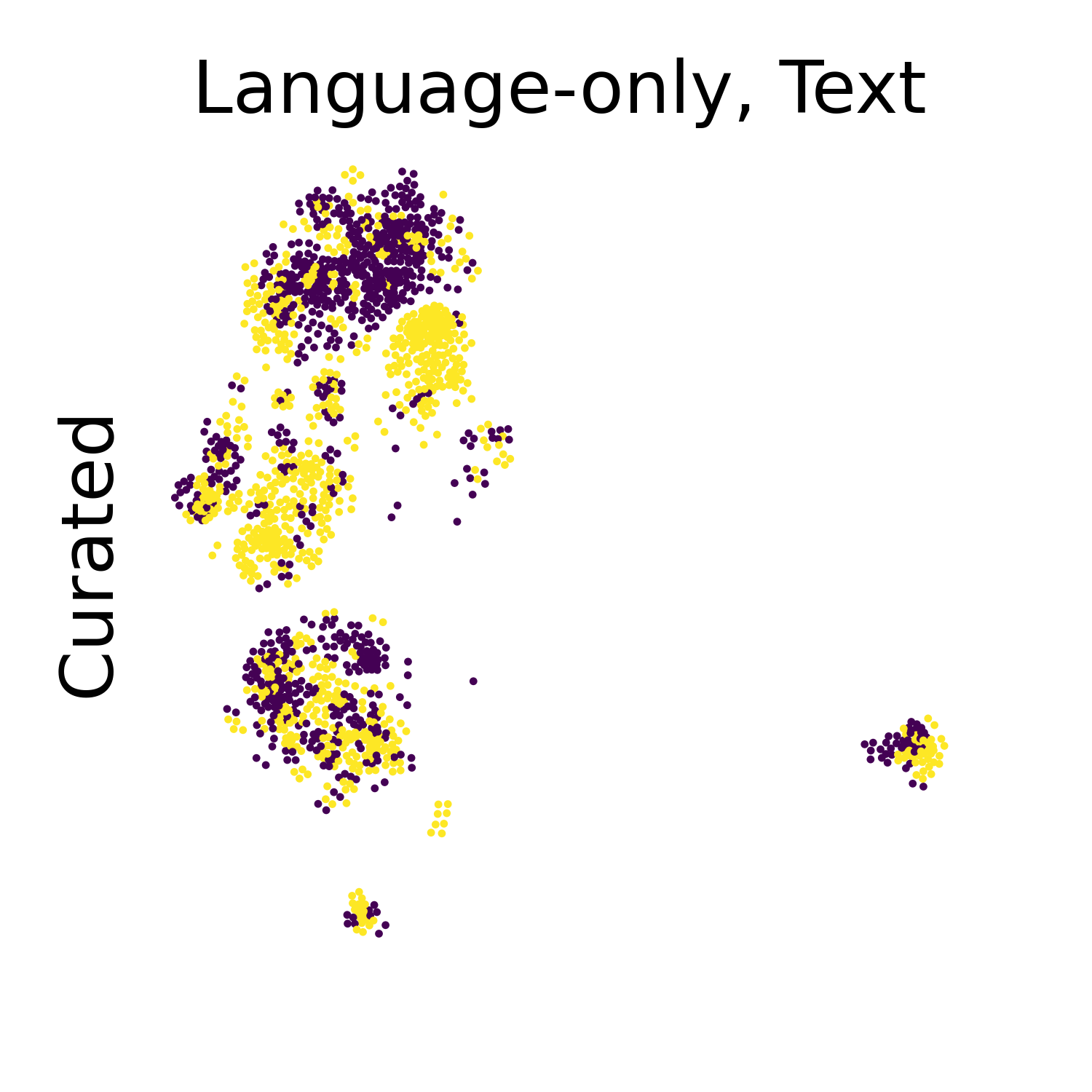}
    \includegraphics[width=0.24\linewidth]{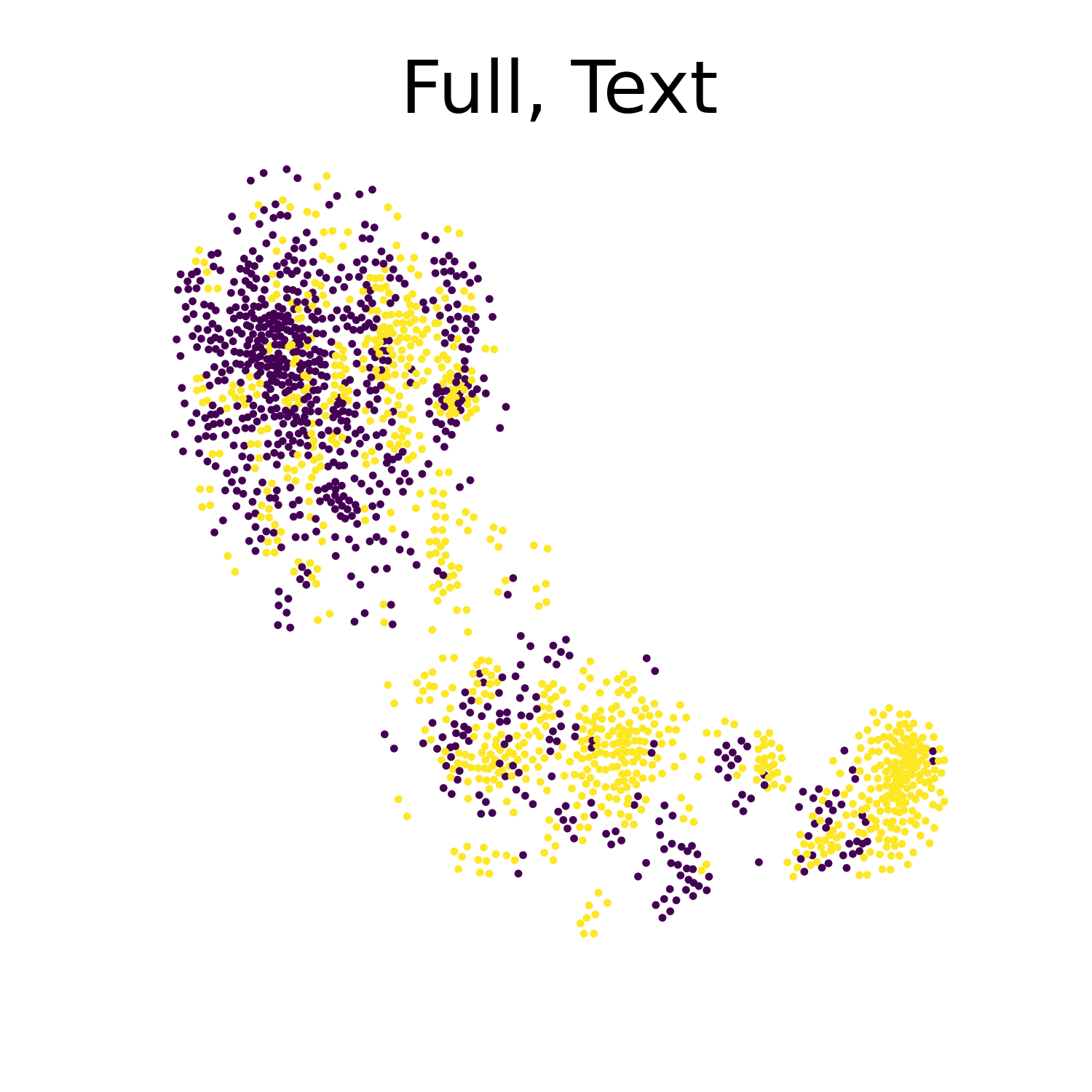}
    \includegraphics[width=0.24\linewidth]{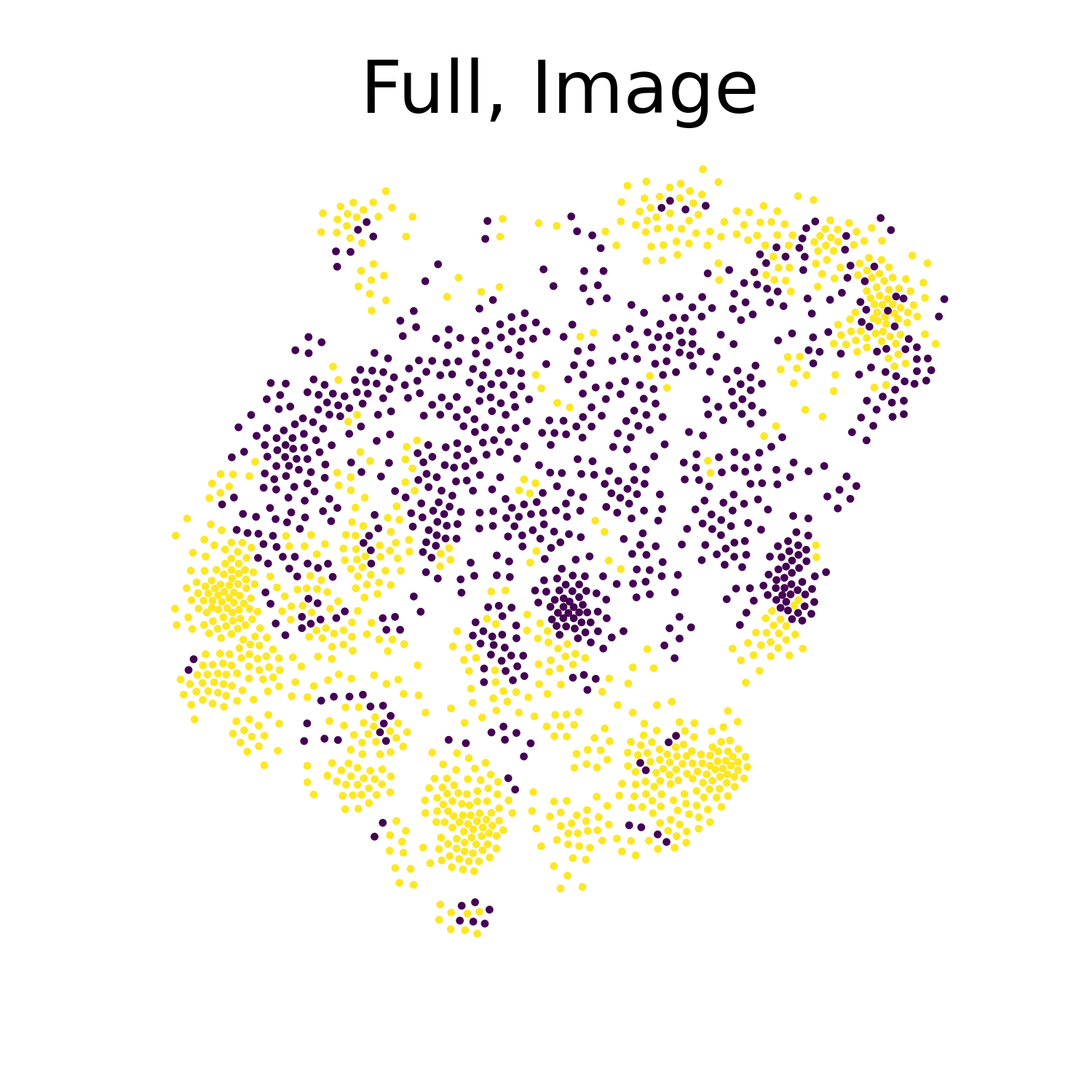}
    \includegraphics[width=0.24\linewidth]{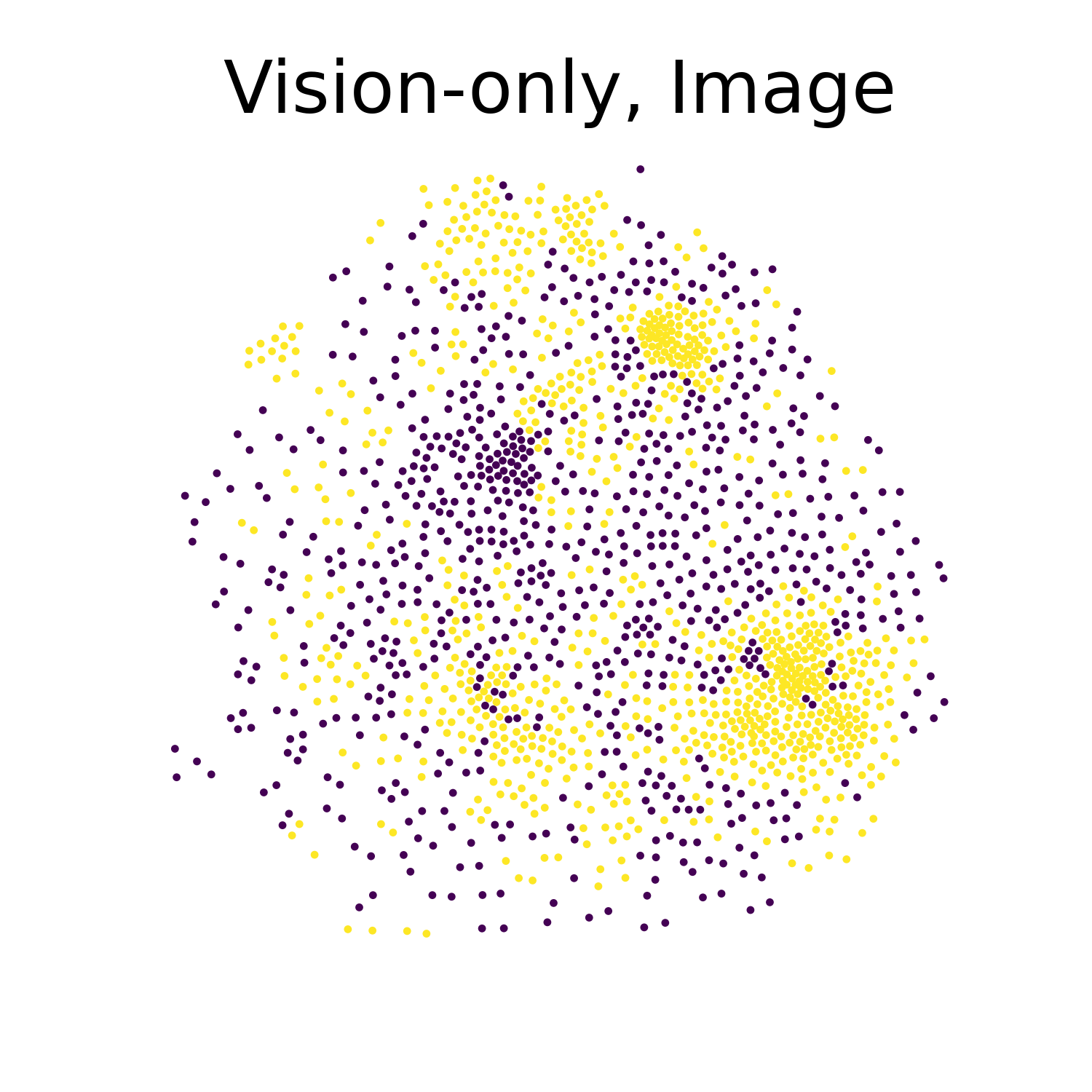}
    \includegraphics[width=0.24\linewidth]{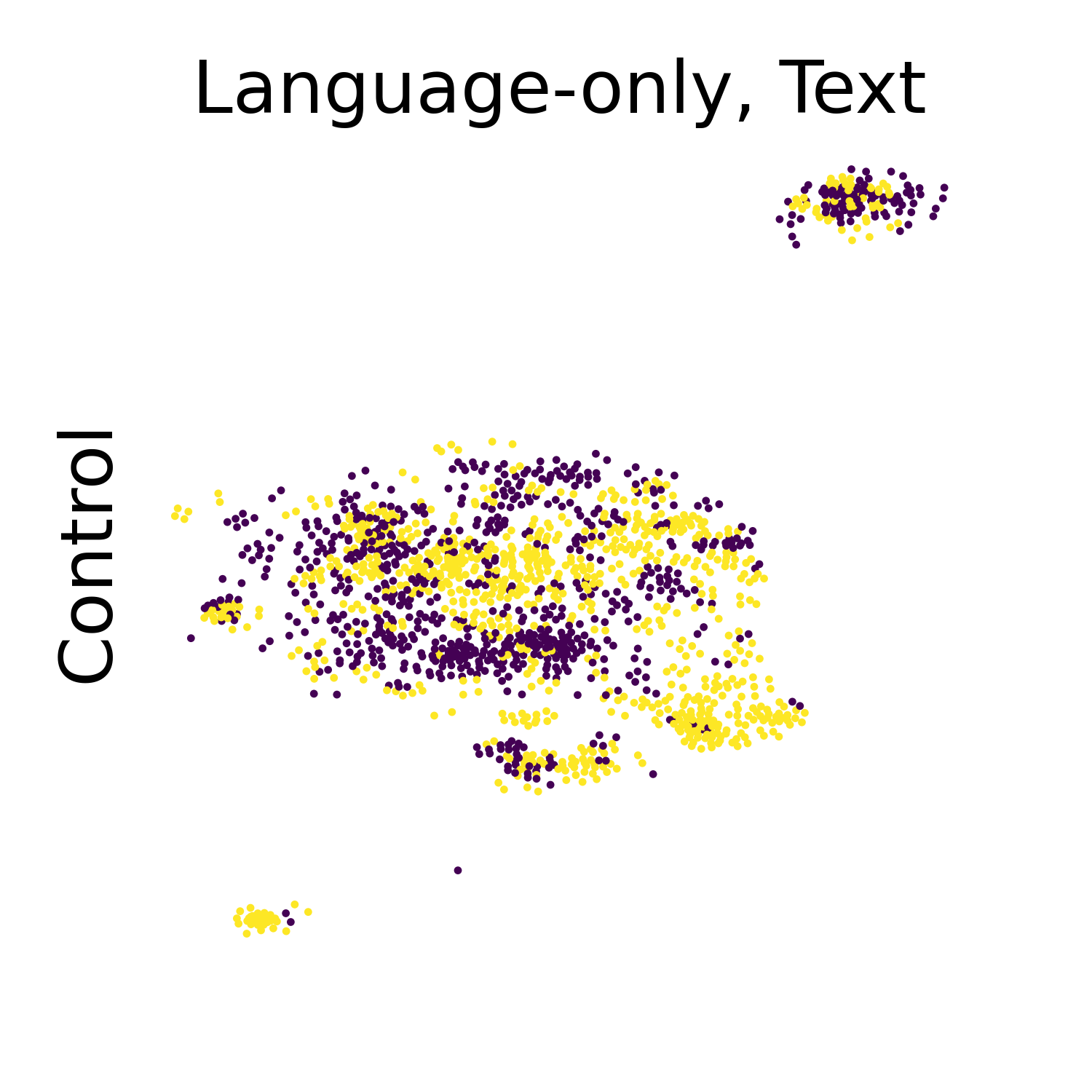}
    \includegraphics[width=0.24\linewidth]{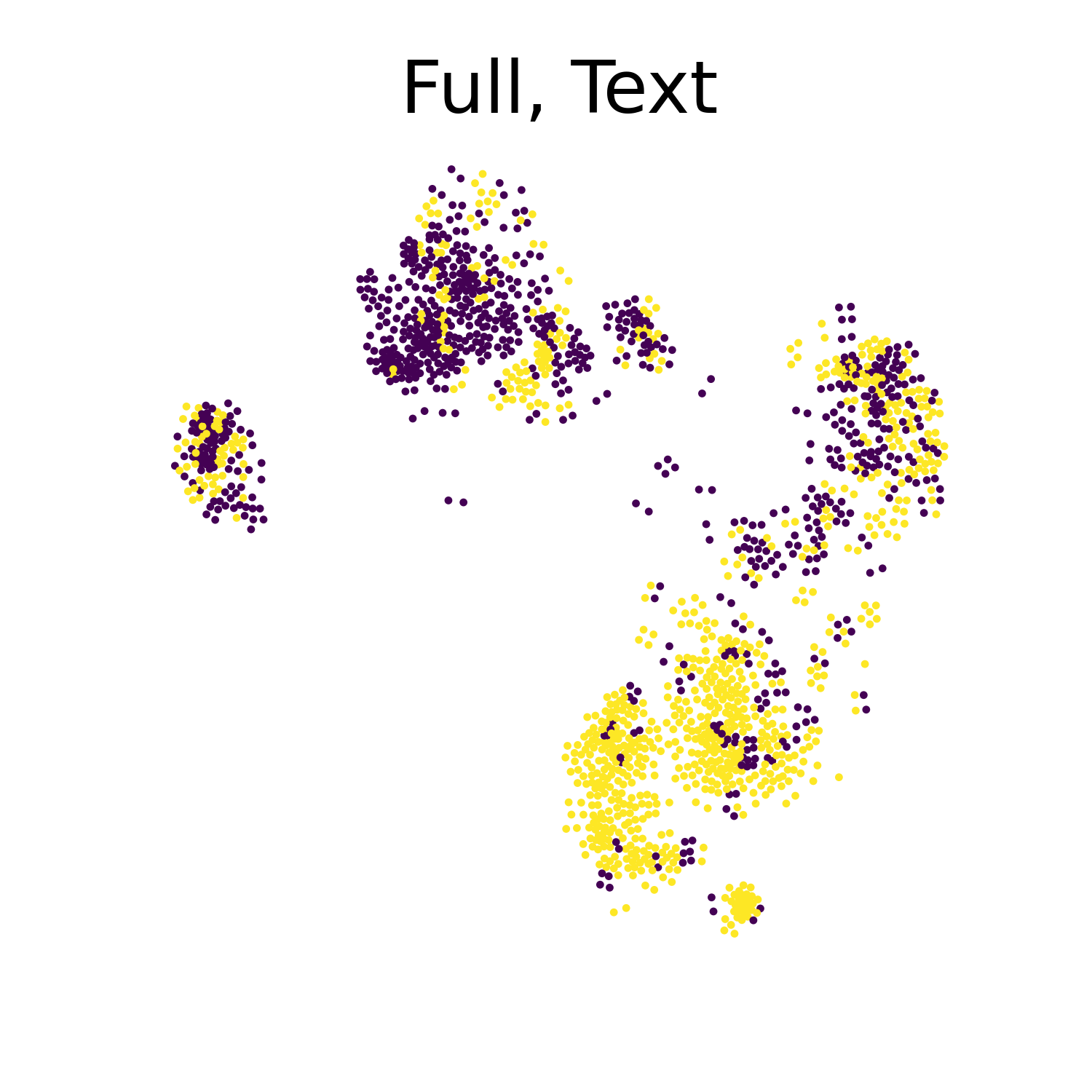}
    \includegraphics[width=0.24\linewidth]{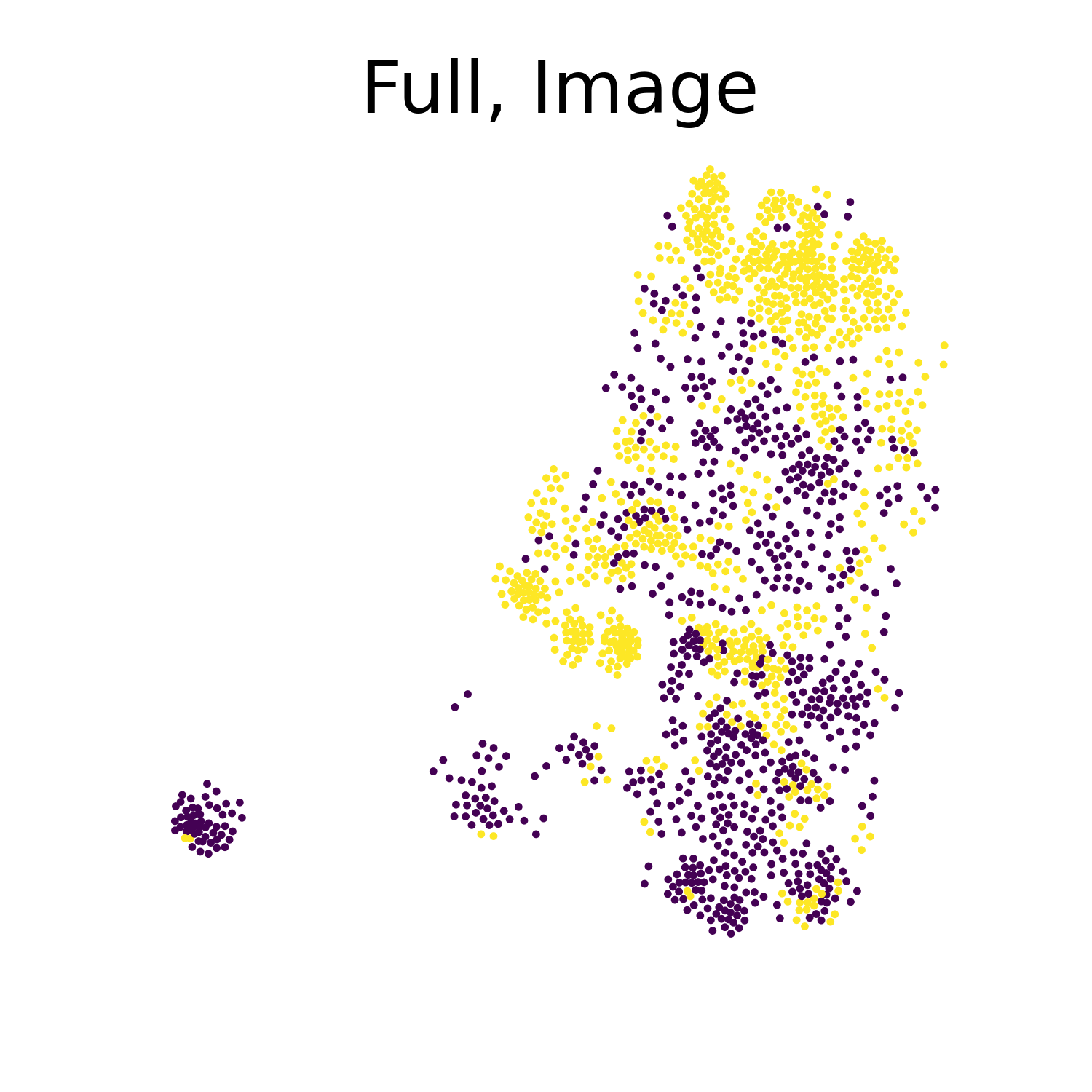}
    \includegraphics[width=0.24\linewidth]{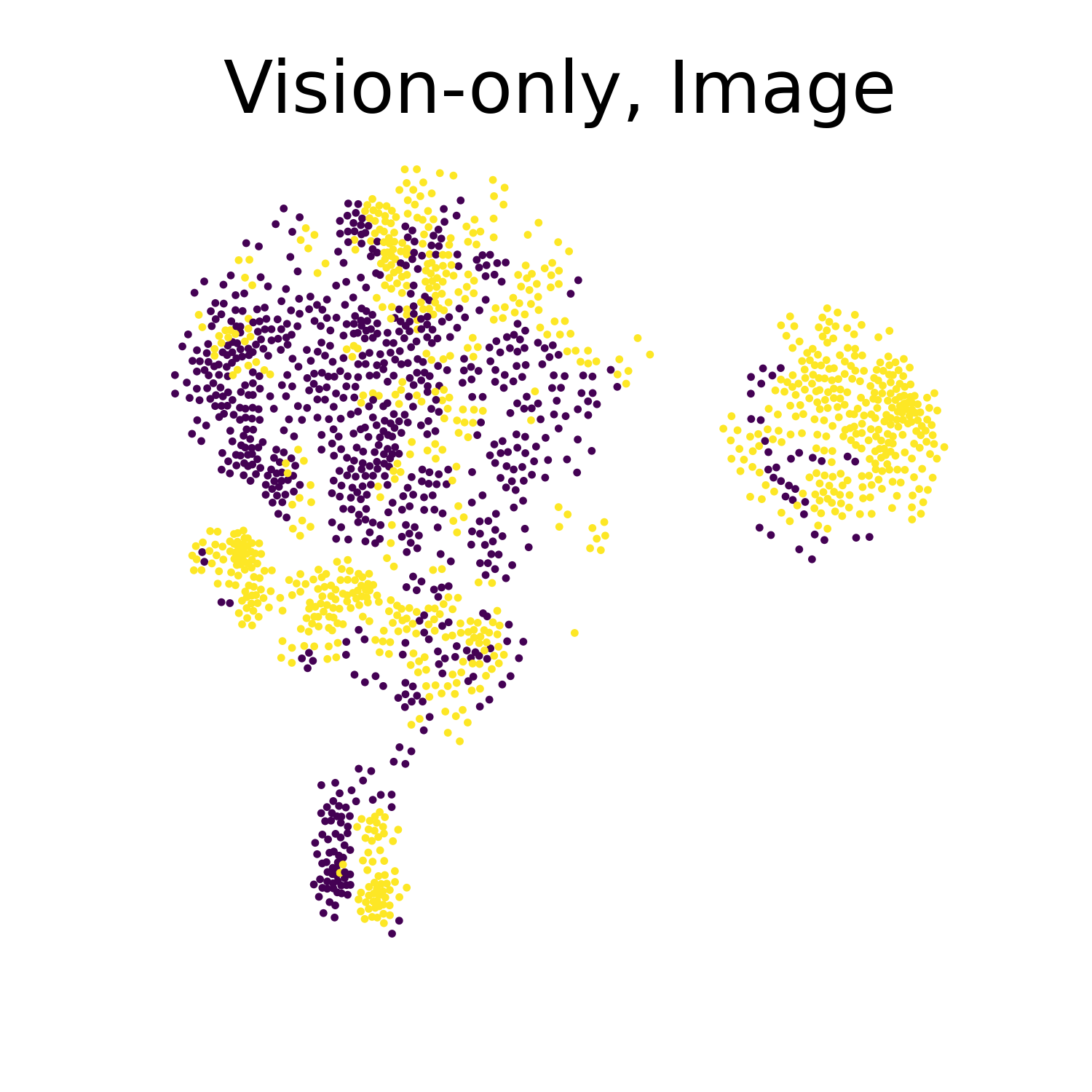}
    \caption{T-SNE visualization of text or image features extracted after various methods of LoRA fine-tuning. The curated caption dataset is shown in the top row, and the control caption dataset in the bottom row. Morphology class FR-I is shown in purple, while FR-II galaxies are labeled in yellow}
    \label{fig:tsne_frozen}
\end{figure}

\end{appendix}

\end{document}